\documentclass[aps,twocolumn,showpacs,prb,superscriptaddress]{revtex4}

\usepackage{dcolumn}
\usepackage{amsmath}
\usepackage{graphicx}

\begin{document}

\title{Planar approximation for the frequencies of spin transfer oscillators}

\author{Ya.\ B. Bazaliy} \email{yar@physics.sc.edu}
 \affiliation{Department of Physics and Astronomy, University of South Carolina,
 Columbia, SC 29208}
 \affiliation{Institute of Magnetism, National Academy of Science, Ukraine.}

\author{F. Arammash}
 \affiliation{Physics and Engineering Department, Benedict College, Columbia, SC 29204}

\date{\today}

\begin{abstract}
A large class of spin transfer oscillators use the free layer with a
strong easy plane anisotropy, which forces its magnetization to move
close to the plane. We show that in this situation the effective
planar approximation provides a fast and accurate way of calculating
the oscillator frequency.
\end{abstract}


\maketitle

\section{Introduction}

Spin transfer devices can have regimes in which their magnetic
moments perform perpetual precessional motions.\cite{slon96,
bjz:2001} In this case they are also called spin torque oscillators
(STO). Magnetic oscillations induced by direct current are
intensively studied experimentally \cite{tsoi:2000, kiselev2003,
rippard:2004, sankey:2005, rippard:2006, houssameddine:2007,
boulle:2007, devolder:2007, houssameddine:2009} and
theoretically.\cite{kent:2004, rezende:2005, bertotti:2005,
serpico:2005:JMMM:oscillations, serpico:2005, bonin:2007, wang2006,
kim:2008:linewidth, tibirkevich:2008, kim:2008:lineshape,
slavin:2008:IEEE} In the STO regime the energy is constantly
supplied to the device from the current source through the spin
transfer mechanism. At the same time it is lost through the usual
dissipation mechanisms, accounted for by the Gilbert damping
constant. In a state of steady precession the energy gain and loss
are balanced on average. In the limit of small damping one observes
the following general picture of the STO
operation.\cite{bertotti:2005} The magnetic moment moves close to
the trajectory which it would follow in the absence of damping and
spin transfer. The actual trajectory is a perturbation of the
zero-damping trajectory, chosen so as to balance the small
dissipation with the equally small energy gain. The main difficulty
in describing the precession states is the lack of knowledge about
the unperturbed trajectory which is a solution of the complicated
non-linear Landau-Lifshitz-Gilbert (LLG) equation. Unless one
considers a small radius precession near an equilibrium point, the
analytic form of such a trajectory is usually unknown and one is
forced to use numeric methods. In this paper we will consider a
special class of spin transfer devices with dominating easy plane
anisotropy. This anisotropy often arises from the thin disk shape of
the magnetic layers found in the majority of experimental
structures. Due to the dominating easy plane anisotropy the LLG
equation can be approximated by an effective planar equation
\cite{bazaliy:2007:APL, bazaliy:2007:PRB, bazaliy:2009:SPIE} that is
less complex and easier to treat analytically.  Here we derive the
planar approximation expressions for the STO oscillation periods and
compare them with the numeric results obtained without
approximations.

\begin{figure}[b]
\center
\includegraphics[width = 0.4\textwidth]{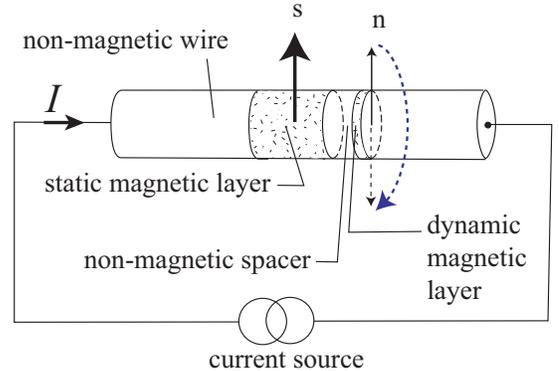}
\caption{Spin transfer device with a fixed and free magnetic layers}
 \label{fig:device}
\end{figure}

\section{Model}

We will consider the case of a spin transfer (Fig.~\ref{fig:device})
device with two magnetic layers, both described as macrospins. One
of the layers has a fixed magnetic moment with a direction given by
a unit vector $\bf s$. This layer acts as a spin polarizer. The
other layers's magnetization is free to move and can be described by
a macrospin magnetic moment ${\bf M}(t) = M_s{\bf n}(t)$ where $M_s$
is the constant saturation magnetization and $\bf n$ is a unit
vector.

The magnetic dynamics of the free layer is governed by the LLG
equation with spin transfer term (see, e.g.,
Ref.~\onlinecite{bjz:2004})
\begin{equation} \label{eq:vector_LLG}
{\dot {\bf n}} = \frac{\gamma}{M_s} \left[ - \frac{\delta E}{\delta
{\bf n}} \times {\bf n} \right] + u [{\bf n} \times [{\bf s} \times
{\bf n}]] + \alpha [{\bf n} \times \dot {\bf n}] \ ,
\end{equation}
where $E$ is the total magnetic energy of the free layer, $\gamma$
is the (positive) gyromagnetic ratio, $\alpha$ is the Gilbert
damping, and the spin-transfer magnitude is given by
\begin{equation}
 \label{eq:u_expression}
u = g(P, ({\vec n} \cdot {\vec s}))\frac{\gamma (\hbar/2)}{V M_s}
\frac{I}{e} \ ,
\end{equation}
where $V$ is the free layer volume, $I$ is the electric current, and
$e$ is the electron charge. The factor $I/e$ is positive when
electrons flow into the free layer. Generally the spin polarization
factor $g(P, ({\vec n} \cdot {\vec s}))$ depends on the angle
between the polarizer and the free layer and on the degree of spin
polarization $P$.\cite{slon96, brouwer, bauer_rmp} In this paper we
will employ the frequently used approximation $g =$ const. In this
case $u$ is simply a rescaled current value.

We consider the free layer with an easy plane anisotropy in the
$(x,y)$ plane and additional easy axis anisotropy along the $\hat x$
direction. External magnetic field $H$ is also applied along $\hat
x$. The magnetic energy is given by
\begin{equation}\label{eq:E}
E({\bf n}) = \frac{K_p}{2} n_z^2 - \frac{K_a}{2} n_x^2 - H M_s n_x
\end{equation}
where $K_p$ and $K_a$ are the easy axis and easy plane anisotropy
constants. To simplify notation we will use the rescaled energy
function
$$
\varepsilon({\bf n}) = \frac{\gamma}{M_s}E = \frac{\omega_p}{2}
n_z^2 - \frac{\omega_a}{2} n_x^2 - h n_x
$$
where the new constants $\omega_p  = \gamma K_p/M_s$, $\omega_a  =
\gamma K_a/M_s$, and $h = \gamma H$ have the dimensions of
frequency.

The fixed layer magnetization is assumed to be pointing along the
easy axis as well, ${\bf s} = + \hat x$.

\section{Effective planar description}

As shown in Ref.~\onlinecite{bazaliy:2007:PRB}, when the
inequalities $\omega_p \gg \omega_a$ and $\omega_p \gg h$ hold,
vector $\bf n$ moves close to the $(x,y)$ plane and its magnetic
dynamics can be described by an effective planar equation governing
the behavior of its in-plane (azimuthal) angle $\phi(t)$, measured
from the $x$-axis. The equation reads
\begin{equation}
 \label{eq:effective_equation}
\frac{\ddot\phi}{\omega_p}  +  \alpha_{eff}(\phi)\ \dot\phi = -
\frac{d \varepsilon_{eff}}{d\phi} \ .
\end{equation}
One can observe that it has a form of a Newton equation of motion
for an ``effective particle'' with a mass $1/\omega_p$ moving in the
external one-dimensional potential $\varepsilon_{eff}(\phi)$ with a
variable friction coefficient $\alpha_{eff}(\phi)$. The effective
energy and friction are found \cite{bazaliy:2007:PRB} to be given by
\begin{eqnarray}
 \nonumber
 \varepsilon_{eff} &=& -\frac{1}{2}\left(\omega_a
 + \frac{u^2}{\omega_p} \right)\cos^2\phi - h\cos\phi  \ ,
 \\
 \label{eq:alpha_DeltaE_special}
 \alpha_{eff} &=& \alpha +
 \frac{2 u \cos\phi}{\omega_p} \ .
\end{eqnarray}
The analogy between the effective planar equation and the Newton
equation for a one-dimensional particle often provides a qualitative
understanding of the system's behavior.

In the absence of spin transfer ($u = 0$) one has $\alpha_{eff} =
\alpha > 0$, and the planar equation predicts that the solution
$\phi(t)$ will approach a minimum of $\varepsilon_{eff}(\phi)$.
Indeed, a particle subjected to a viscous friction force will
eventually come to rest in one of the energy minima. This is
guaranteed by a classical mechanics theorem, stating that the total
energy
\begin{equation}\label{total_energy}
\varepsilon_{tot}(t) = \frac{\dot\phi^2}{2\omega_p} +
\varepsilon_{eff}(\phi)
\end{equation}
is a decreasing function of time. The theorem is based on the
relationship
\begin{equation}\label{total_energy_derivative}
\frac{d\varepsilon_{tot}}{dt} = - \alpha_{eff}(\phi) \dot\phi^2 \ ,
\end{equation}
and holds for $\alpha_{eff} > 0$ (case of conventional friction).

When the current is turned on, the effective friction can become
negative for some values of $\phi$. In the areas of negative
friction the energy of the effective particle may increase, the
theorem about the decrease of $\varepsilon_{tot}(t)$ breaks down,
and the particle motion does not necessarily have to stop in the
local energy minimum. As a result, the effective particle can
perform an oscillating motion which corresponds to the persistent
precession state of the free layer.

\subsection{Description of oscillation regimes}
\label{sec:oscillation_regimes}

\begin{figure}[b]
    \resizebox{.45\textwidth}{!}{\includegraphics{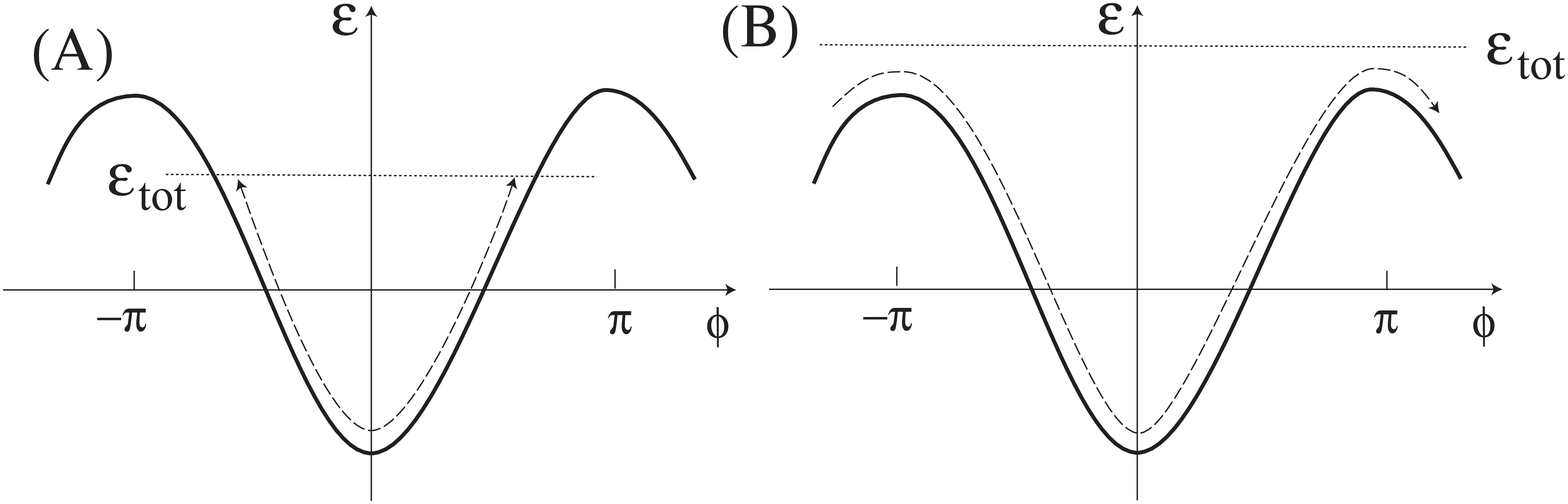}}
\caption{Oscillation states at $h > \tilde\omega$. The solid line
gives the effective energy $\varepsilon_{eff}(\phi)$. Dashed lines
with arrows show the motion of the particle in the potential
profile. Horizontal dotted line gives the level of the total energy
$\varepsilon_{tot}$. (A) small oscillations at $-u_2 < u < -u_1$.
(B) full rotations at $u < -u_2$.}
 \label{fig:oscillations_large_h}
\end{figure}

In the case of effective energy and friction given by
Eqs.~\ref{eq:alpha_DeltaE_special} the effect of $u \neq 0$  was
analyzed in Ref.~\onlinecite{bazaliy:2007:PRB}. We start with a
review of these results. Denoting $\tilde \omega = \omega_a +
u^2/\omega_p$, one finds that the effective energy has a minimum at
$\phi = 0$ for $h
> -\tilde\omega$ and $\phi = \pi$ for $h < \tilde\omega$. Both
minima are present for $|h| < \tilde\omega$. For small currents the
effective damping is still positive for all values of $\phi$. As the
current increases, $\alpha_{eff}$ first becomes negative at one of
the minimum points $\phi = 0$ or $\phi = \pi$ at the critical
current values $u = \mp u_1$, where
$$
u_1 = \frac{\alpha\omega_p}{2} \ .
$$
Negative $\alpha_{eff}$ at a point of energy minimum leads to the
development of oscillations around this formerly stable equilibrium.
As the current grows, the region of negative $\alpha_{eff}$ becomes
larger and the oscillations amplitude increases. The evolution of
oscillations at large amplitudes proceeds differently for $|h| <
\tilde\omega$ and $|h| > \tilde\omega$.

In the large field case, $|h| > \tilde\omega$, the energy has only
one minimum. For definiteness, consider the case of positive $h >
\tilde\omega$, with the minimum of $\varepsilon_{eff}$ at $\phi = 0$
and a maximum at $\phi = \pi$ (Fig.~\ref{fig:oscillations_large_h}).
The effective damping at the minimum point changes sign to negative
at $u = - u_1$. After that the system develops small oscillations
around $\phi = 0$ (Fig.~\ref{fig:oscillations_large_h}A). As the
current is made even more negative, the amplitude of these
oscillations grows, until at a second critical current $u = -u_2(h)$
it becomes so large that the effective particle reaches the point of
energy maximum $\phi = \pi$. After that the particle starts to make
full circles in the easy plane
(Fig.~\ref{fig:oscillations_large_h}B). The speed of the particle
performing full circles becomes larger and larger as the current is
decreased further below $-u_2$.

\begin{figure}[t]
    \resizebox{.45\textwidth}{!}{\includegraphics{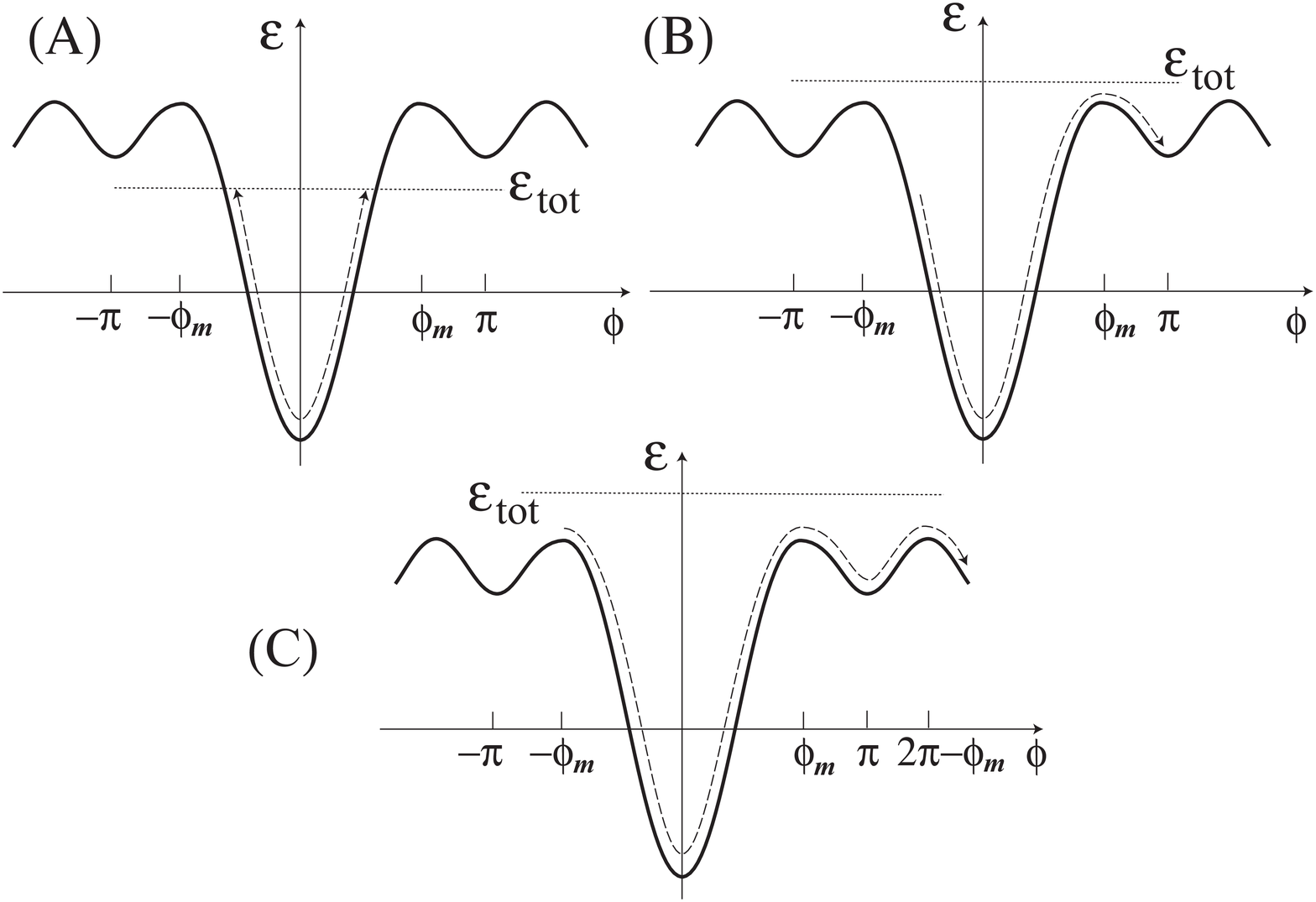}}
\caption{Oscillation states and switching at $0 < h < \tilde\omega$.
The solid line gives the effective energy $\varepsilon_{eff}(\phi)$.
Dashed lines with arrows show the motion of the particle in the
potential profile. Horizontal dotted line gives the level of the
total energy $\varepsilon_{tot}$. (A) small oscillations at $-u_2 <
u < -u_1$; (B) Switching to the $\phi = \pi$ local minimum at $-u_3
< u < -u_2$; (C) Full rotations at $u < -u_3$.}
 \label{fig:oscillations_small_h}
\end{figure}

\begin{figure}[t]
    \resizebox{.45\textwidth}{!}{\includegraphics{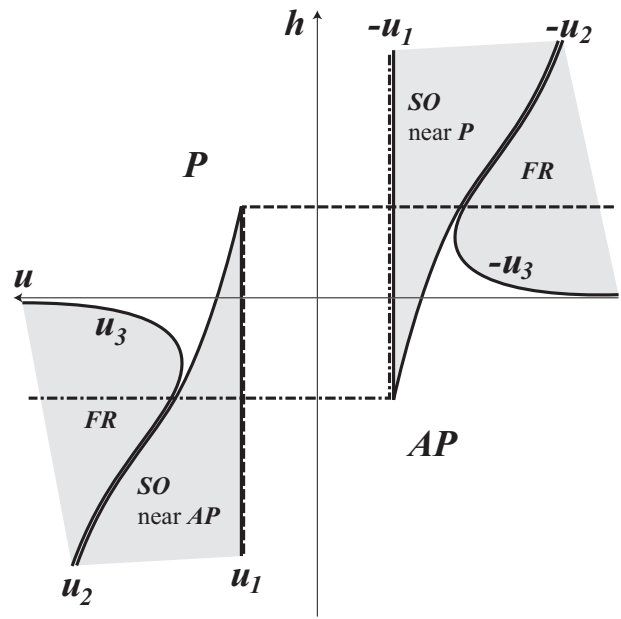}}
\caption{Switching diagram. The dashed and dash-dotted lines give
the stability boundaries for the $\phi = \pi$, antiparallel (AP) and
$\phi = 0$, parallel (P) states respectively. These static states
coexist with each other in the central rectangle of the diagram. The
oscillatory regions are shown as grey areas, marked by the type of
oscillation: small oscillations (SO) near P or AP points or full
rotation regime (FR). Oscillations sometimes coexist with P or AP
states. The diagram is symmetric in accord with the condition
(\ref{eq:diagram_symmetry}).}
 \label{fig:switching_diagram}
\end{figure}

In the small field case, $|h| < \tilde\omega$, there are two minima
of $\varepsilon_{eff}$ at $\phi = 0$ and $\phi = \pi$ with a maximum
point at the angle $\phi_{m}(h)$ between them
(Fig.~\ref{fig:oscillations_small_h}). Considering again the case of
positive field $0 < h < \tilde\omega$ one finds the following
picture. The $\phi = 0$ minimum is destabilized at a negative
current $u = -u_1$ and small oscillations around this equilibrium
are developed (Fig.~\ref{fig:oscillations_small_h}A). As the current
is decreased further, the amplitude of the oscillations grows until
they reach the point of energy maximum at the second critical
current $u = -u_2(h)$. For larger amplitudes the effective particle
moves over the energy maximum into the basin of the $\phi = \pi$
minimum (Fig.~\ref{fig:oscillations_small_h}B). In this basin the
effective damping is positive and the particle gradually looses its
total energy, ending up in the $\phi = \pi$ minimum.

There is, however, yet another transition at the third critical
current $u = -u_3(h)$. Beyond the third threshold the particle
reaches the $\phi = \phi_m$ maximum with a velocity that is
sufficiently high to allow it to move all the way to the next energy
maximum at $\phi = 2\pi-\phi_m$, overcoming the friction force that
attempts to stop it (Fig.~\ref{fig:oscillations_small_h}C). In other
words, the energy obtained in the negative damping area around $\phi
= 0$ is sufficient to push the particle around the full circle from
$-\phi_m$ to $2\pi - \phi_m$. At $u < - u_3(h)$ the particle can be
either in the oscillation regime with full circle rotations, or at
rest in the $\phi = \pi$ energy minimum. The actual state is
determined by the history of the system.

So far we have discussed small oscillations near the $\phi = 0$
equilibrium and their transformation into the full rotation regime.
Positive current $u > 0$ can destabilize the $\phi = \pi$
equilibrium. Due to the symmetry of the problem, the critical
currents $u_i$ of the $\phi = \pi$ equilibrium are related to those
already found for $\phi = 0$ as
\begin{equation}\label{eq:diagram_symmetry}
u_i(\pi,h) = - u_i(0,-h) \qquad i = 1,2,3
\end{equation}

The switching diagram\cite{bazaliy:2007:PRB} of our spin transfer
system is shown in Fig.~\ref{fig:switching_diagram}. It consists of
the two 90-degree wedges of stability of the $\phi = 0$ equilibrium
(``parallel state''), and $\phi = \pi$ equilibrium (``antiparallel
state''). In addition to the stability areas of the fixed points
there are stability areas of the oscillating solutions (grey areas
in Fig.~\ref{fig:switching_diagram}). The oscillating solutions of
the planar equation are in a one-to-one correspondence with the free
layer precession states previously found numerically by solving the
LLG equation.\cite{kiselev2003, xiao2005} The small oscillations
regime corresponds to the case of in-plane precession of vector $\bf
n$, and the full rotation regime corresponds to the out-of-plane
precession.\cite{bazaliy:2007:PRB}

As the current is increased deeper into the full rotation regime the
deviations of $\bf n$ from the $(x,y)$ easy plane plane grow, and
the planar approximation becomes inapplicable. The orbits of
persistent precession evolve \cite{bertotti:2005, xiao2005,
slavin:2008:IEEE} into small circles around the maximum of the
magnetic energy (\ref{eq:E}) far away from the easy plane.
Eventually this point is stabilized by spin
transfer.\cite{bjz:2001,bjz:2004}

\subsection{Small damping approximation}\label{sec:small_damping}
Consider first the planar equation (\ref{eq:effective_equation}) at
$u = 0$. In the limit of $\alpha \to 0$ the friction term can be
viewed as a perturbation on top of the frictionless motion. When
friction is completely absent, the total energy is exactly conserved
\begin{equation}\label{total_energy_conservation}
\varepsilon_{tot}(t) = \varepsilon_{tot}(0)
\end{equation}
For small friction $\varepsilon_{tot}$ is conserved approximately,
i.e., its relative change during one period of oscillations around
the potential minimum is small. On a time scale of several periods
one can approximately use equation
(\ref{total_energy_conservation}). For longer time intervals one has
to take into account that $\varepsilon_{tot}(t)$ is a slowly
changing function of time.

In the case of $u \neq 0$ the same picture holds as long as the
absolute value of $\alpha_{eff}$ is small. The only difference is
that now the total energy can either decrease or grow during one
period of oscillations. Here we want to note that the smallness of
$|\alpha_{eff}|$ is already guaranteed by the smallness of $\alpha$
in the following sense. As discussed above, the oscillating regimes
of spin transfer systems appear when the effective damping becomes
negative. At the onset of the oscillation regime the Gilbert damping
term and the spin transfer term in the expression for $\alpha_{eff}$
are of the same order. That means that $|\alpha_{eff}|$ and $\alpha$
have the same order of magnitude, unless the current is increased
far above the threshold.

The slow time change of $\varepsilon_{tot}$ can be calculated using
the following procedure. We use Eq.~(\ref{total_energy_derivative})
to find the energy change during one period of oscillations
$$
\Delta = -\int_0^T \alpha_{eff} \dot\phi^2 dt \ .
$$
Applying the approximate energy conservation during the period one
can express the particle velocity as
\begin{equation}\label{eq:approximate_dot_phi}
\dot\phi \approx \pm \sqrt{2 \omega_p (\varepsilon_{tot} -
\varepsilon_{eff}(\phi))} = f(\phi,\varepsilon_{tot})
\end{equation}
where $\varepsilon_{tot}$ is assumed to be constant, for example
taken as the value of the total energy at the beginning of the
period. The plus or minus sign in front of the expression is chosen
according to the direction of particle motion. The electric current
parameter $u$ enters $\alpha_{eff}$ as a linear correction, and
$\varepsilon_{eff}$ as a quadratic correction. It will be shown in
Sec.~\ref{sec:conditions} that the correction to the effective
energy can be dropped in our approximation and $\varepsilon_{eff}$
can be substituted by $\varepsilon(\phi) = - (1/2)\omega_a\cos^2\phi
- h\cos\phi$ in Eq.~(\ref{eq:approximate_dot_phi}), i.e., one can
use $\omega_a$ instead of $\tilde\omega$.

Using approximation (\ref{eq:approximate_dot_phi}) one finds both
$\Delta$ and the oscillation period $T$ as
\begin{eqnarray} \label{eq:integrals_for_DeltaE_T}
\Delta  &\approx& - \oint \alpha_{eff}(\phi) f(\phi) \ d\phi \ ,
 \\
 \nonumber
T &=& \int_0^T dt = \oint \frac{d\phi}{\dot\phi} \approx \oint
\frac{d\phi}{f(\phi)} \ .
\end{eqnarray}
In these formulae the integrals over $\phi$ are taken along the
closed trajectory corresponding to one period of oscillations in the
absence of damping and spin transfer. The value of
$\varepsilon_{tot}$ determines both the integration limits, and the
integrand. Consequently, equations (\ref{eq:integrals_for_DeltaE_T})
give both $\Delta$ and $T$ as the functions of $\varepsilon_{tot}$.
The slow time evolution of the total energy is described by an
approximate equation
\begin{equation}\label{eq:totenergy_diff_equation}
\frac{d\varepsilon_{tot}}{dt} \approx
\frac{\Delta(\varepsilon_{tot})}{T(\varepsilon_{tot})}
\end{equation}

In the regime of persistent oscillations the total energy is
constant, which implies $\Delta(\varepsilon_{tot}) = 0$. In
Ref.~\onlinecite{bazaliy:2007:PRB} this condition was used to find
the critical currents $u_i$ by the following argument. The critical
trajectories corresponding to each $u_i$ were already described in
Sec.~\ref{sec:oscillation_regimes}. On those trajectories the value
of $\varepsilon_{tot}$ is given by the value of $\varepsilon_{eff}$
at the turning points where $\dot\phi = 0$. Now the integral
(\ref{eq:integrals_for_DeltaE_T}) can be calculated as a function of
$u$, and the equation $\Delta(u) = 0$ gives a current threshold.

\subsection{Oscillation periods}
Here we use (\ref{eq:totenergy_diff_equation}) to calculate the
period of oscillations with arbitrary amplitude. Our goal is to find
the function $T(u)$. The equation $d\varepsilon_{tot}/dt = 0$ or
$\Delta = 0$ holds for persistent oscillation regimes at any
current, not just for the critical current values. At a given $u$
this equation determines the value of $\varepsilon_{tot}$ and the
endpoints of the integration contour in the first equation of the
system (\ref{eq:integrals_for_DeltaE_T}). Knowing them, we can
calculate $T$ from the second equation in
(\ref{eq:integrals_for_DeltaE_T}) and find the function $T(u)$.

We will do the calculations for $u < 0$. According to the symmetry
of the switching diagram (Fig.~\ref{fig:switching_diagram}) the $u >
0$ results can be easily obtained from those for $u < 0$.

\subsubsection{Case of $h > \omega_a$}

Here one observes small oscillations for $-u_2 < u < -u_1$ and full
rotations for $u < -u_2$.

In the small oscillations regime the particle moves between the
points of maximum deviation $\pm \phi_a$
(Fig.~\ref{fig:oscillations_large_h}A). The total energy is given by
$\varepsilon_{tot} = \varepsilon(\phi_a) = (1/2)\omega_a
\cos^2\phi_a - h \cos\phi_a$, and the expression
(\ref{eq:approximate_dot_phi}) specializes to
\begin{eqnarray*}
 \dot\phi & \approx & \pm \Omega \sqrt{R(\phi)} \ ,
 \\
 R(\phi,\phi_a,h) &=&
 \cos^2\phi - \cos^2\phi_a +
 \frac{2 h}{\omega_a}(\cos\phi-\cos\phi_a) \ ,
 \\
 \Omega &=& \sqrt{\omega_a \omega_p} \ .
\end{eqnarray*}
Condition $\Delta = 0$ together with
(\ref{eq:integrals_for_DeltaE_T}) gives
$$
\int_{-\phi_a}^{\phi_a}
  \left(\alpha + \frac{2u\cos\phi}{\omega_p}\right)
  \Omega \sqrt{R(\phi)}  \ d\phi
  = 0
$$
or
$$
\alpha K_1(\phi_a) + \frac{2u}{\omega_p} K_2(\phi_a) = 0
$$
with
\begin{eqnarray*}
&& K_1(\phi_a) = \int_{-\phi_a}^{\phi_a} \sqrt{R(\phi)} \ d\phi \ ,
 \\
&& K_2(\phi_a) = \int_{-\phi_a}^{\phi_a} \cos\phi \sqrt{R(\phi)} \
d\phi \ .
\end{eqnarray*}
We can now express the current as a function of the oscillation
amplitude $\phi_a$
\begin{equation}\label{eq:u_vs_phia_at_large_h}
u(\phi_a) = -\frac{\alpha\omega_p}{2}
\frac{K_1(\phi_a)}{K_2(\phi_a)} \ .
\end{equation}
The second equation of (\ref{eq:integrals_for_DeltaE_T}) gives the
oscillation period, also expressing it as a function of $\phi_a$
\begin{equation}\label{eq:T_vs_phia_at_large_h}
T(\phi_a) = 2 \int_{-\phi_a}^{\phi_a}
\frac{d\phi}{\Omega\sqrt{R(\phi)}} = \frac{2 K_3(\phi_a)}{\Omega} \
.
\end{equation}
Using equations (\ref{eq:u_vs_phia_at_large_h}) and
(\ref{eq:T_vs_phia_at_large_h}) one can make a parametric plot
$T(u)$ by varying $\phi_a$ from zero to $\pi$. The integrals $K_i$
can be expressed through special functions or, in practice,
calculated by a computer algebra system.

The boundary between the small oscillations and the full rotations
is found at the current value
\begin{equation}\label{eq:u2_large_h}
-u_2(h) = u(\pi) = -\frac{\alpha\omega_p}{2}
\frac{K_1(\pi,h)}{K_2(\pi,h)}  \quad (h > \omega_a) \ ,
\end{equation}
where we have explicitly indicated the dependence of the integrals
on $h$. Note that this formula corrects formula (8) from
Ref.~\onlinecite{bazaliy:2007:PRB} which is valid only for the
values of $h$ slightly above $\omega_a$, i.e., for $h - \omega_a \ll
\omega_a$.

Below the $-u_2$ threshold the oscillations happen with a full
rotation of the angle $\phi$. There are no endpoints here
(Fig.~\ref{fig:oscillations_large_h}B) and we will parameterize the
trajectory by the excess energy $\Delta\varepsilon =
\varepsilon_{tot} - \varepsilon(\pi) > 0$. Equation
(\ref{eq:approximate_dot_phi}) specializes to
$$
\dot\phi = \Omega \sqrt{R(\phi,\pi,h) +
\frac{2\omega_p\Delta\varepsilon}{\Omega}}
 = \Omega \sqrt{R(\phi,\pi,h) + \delta}
$$
where we have introduced $\delta =
2\omega_p\Delta\varepsilon/\Omega$. Repeating the steps that led to
equations (\ref{eq:u_vs_phia_at_large_h}) and
(\ref{eq:T_vs_phia_at_large_h}) we get an analogous pair of
equations for the full rotation regime
\begin{eqnarray}
 \label{eq:u_vs_de_at_large_h} u(\delta) &=&
-\frac{\alpha\omega_p}{2} \frac{L_1(\delta)}{L_2(\delta)}
 \\
 \label{eq:T_vs_de_at_large_h} T(\delta) &=&
 \frac{L_3(\delta)}{\Omega}
\end{eqnarray}
where the integrals $L_i$ are expressed through $\tilde R =
R(\phi,\pi,h)$ as
\begin{eqnarray*}
&& L_1(\delta) = \int_{-\pi}^{\pi}
 \sqrt{\tilde R(\phi)+\delta} \ d\phi
 \\
&& L_2(\delta) = \int_{-\pi}^{\pi} \cos\phi
 \sqrt{\tilde R(\phi)+\delta} \ d\phi
 \\
&& L_3(\delta) = \int_{-\pi}^{\pi}
 \frac{d\phi}{\sqrt{\tilde R(\phi) + \delta}}
\end{eqnarray*}
Using equations (\ref{eq:u_vs_de_at_large_h}) and
(\ref{eq:T_vs_de_at_large_h}) one can make a parametric plot $T(u)$
by varying $\delta$ from zero up.

\subsubsection{Case of $|h| < \omega_a$}

For small oscillations ($-u_2 < u < -u_1$), the calculations turn
out to be identical with those performed in the previous section.
Expressions (\ref{eq:u_vs_phia_at_large_h}) and
(\ref{eq:T_vs_phia_at_large_h}) can be used to make a parametric
plot $T(u)$. The only difference is that now the oscillation
amplitude $\phi_a$ changes form zero to the angle $\phi_m$ of the
energy maximum.

The critical current $u_2(h)$ is found by setting $\phi_a = \phi_m$.
The position of the energy maximum is found from the equation
$d\varepsilon/d\phi = 0$ which gives
$$
\cos\phi_m = -\frac{h}{\omega_a} \ .
$$
At the same time the expression for $R$ can be rewritten as
$$
R = \left( \cos\phi + \frac{h}{\omega_a} \right)^2 -
  \left( \cos\phi_a + \frac{h}{\omega_a} \right)^2
$$
As a result, for $\phi_a = \phi_m$ one is able to write an explicit
formula for the square root
\begin{equation}\label{eq:sqrtR_at_phim}
\sqrt{R(\phi,\phi_m,h)} = |\cos\phi + h/\omega_a| \ .
\end{equation}
The integrals $K_{1,2}$ can be then taken and one gets the
expression for the critical current\cite{bazaliy:2007:PRB}
\begin{equation}\label{eq:u2_small_h}
u_2 = \alpha\omega_p
 \frac{\sin\phi_m + (h/\omega_a) \phi_m}{\phi_m + (h/\omega_a)
 \sin\phi_m},
 \quad (|h| < \omega_a) \ .
\end{equation}

The third critical current $u = -u_3(h)$ corresponds to the
trajectory shown in Fig.~\ref{fig:oscillations_small_h}C. The angle
changes from $-\phi_m$ to $2\pi - \phi_m$, and the total energy is
equal to $\varepsilon(2\pi - \phi_m)$ because at the critical
current the particle starts at $-\phi_m$ with zero velocity, makes a
full rotation, and reaches $2\pi - \phi_m$ with zero velocity. Since
$\varepsilon(2\pi - \phi_m) = \varepsilon(\phi_m)$, we find that the
condition $\Delta = 0$ specializes to
$$
\int_{-\phi_m}^{2\pi - \phi_a}
  \left(\alpha + \frac{2u\cos\phi}{\omega_p}\right)
  \sqrt{R(\phi,\phi_m,h)}  \ d\phi
  = 0
$$
(the integrand is the same as in the case of small oscillations but
the integration limits are different). Using equation
(\ref{eq:sqrtR_at_phim}) we then find\cite{bazaliy:2007:PRB}
\begin{equation}\label{eq:u3_small_h}
u_3 = \alpha\omega_p
 \frac{\sin\phi_m + (h/\omega_a) (\phi_m - \pi/2)}{(\phi_m - \pi/2) +
 (h/\omega_a)
 \sin\phi_m},
 \quad (|h| < \omega_a) \ .
\end{equation}

For the currents exceeding the third threshold, $u < - u_3$, there
are no turning points and, similar to the case of large fields, we
parameterize the trajectory by the scaled energy excess over the
maximum: $\delta = 2\omega_p (\varepsilon(\phi) -
\varepsilon(\phi_m))/\Omega
> 0$. The resulting parametric expressions for the current and
period turn out to be identical to
Eqs.~(\ref{eq:u_vs_de_at_large_h}) and
(\ref{eq:T_vs_de_at_large_h}), except that the meaning of $\tilde R$
changes to $\tilde R = R(\phi,\phi_m,h) = (\cos\phi + h/\omega_a)^2$
in the definitions of the integrals $L_{1,2,3}$.

\begin{figure}[t]
    \resizebox{.45\textwidth}{!}{\includegraphics{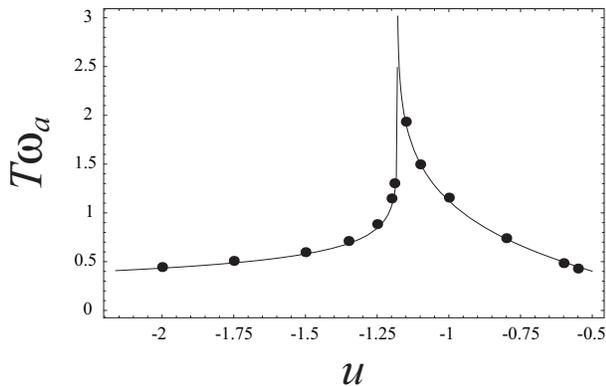}}
\caption{Large field case. Comparison between the no-approximation
numeric results (dots) and planar approximation expressions (lines)
for the oscillations periods. Here $\omega_p = 100 \ \omega_a$, $h =
1.4 \ \omega_a$, $\alpha = 0.01$. The oscillation period diverges at
$u = -u_2$}
 \label{fig:compare_large_h}
\end{figure}

\begin{figure}[t]
    \resizebox{.4\textwidth}{!}{\includegraphics{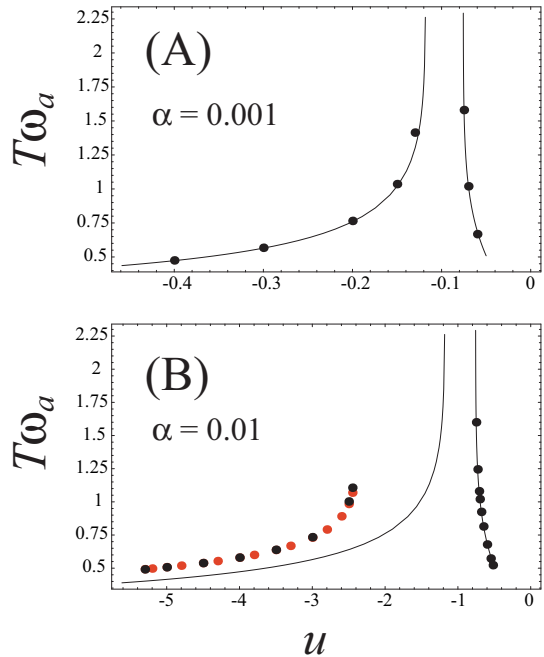}}
\caption{(Color online) Small field case. Comparison between the
no-approximation numeric results (dots) and planar approximation
expressions (line) for the oscillations periods for $\omega_p = 100
\ \omega_a$, $h = 0.5 \ \omega_a$. The oscillation period diverges
at $u = -u_2$ and $u = -u_3$ with no oscillations between the two
thresholds. Top panel (A) gives results for $\alpha = 0.001$, where
the agreement is very good. The bottom panel (B) shows a comparison
for $\alpha = 0.01$. Planar approximation works well for small
precession, but the full rotations regime requires currents which
are too large for the small friction approximation
(\ref{eq:u_vs_de_at_large_h}, \ref{eq:T_vs_de_at_large_h}) to be
accurate. The exact solution of the planar equation (red (grey)
dots) is still in very good agreement with the numeric solution of
the LLG equation (black dots).}
 \label{fig:compare_small_h}
\end{figure}

Overall, equation pairs (\ref{eq:u_vs_phia_at_large_h},
\ref{eq:T_vs_phia_at_large_h})  and (\ref{eq:u_vs_de_at_large_h},
\ref{eq:T_vs_de_at_large_h}) give the planar approximation formulae
for the periods of all spin transfer oscillations possible in our
system. For small oscillations regime with current just above the
first threshold the results (\ref{eq:T_vs_phia_at_large_h}) and
(\ref{eq:T_vs_de_at_large_h}) give the frequencies converging to
$\sqrt{(\omega_a + h)\omega_p}$, i.e., to the Kittel's formula with
a substitution $\omega_p + h \to \omega_p$ in accord with our
approximation $h \ll \omega_p$. For other current values we will
present the results of the theory as graphs
(Figs.~\ref{fig:compare_large_h} and \ref{fig:compare_small_h}) for
the periods $T(u,h)$ rather than for the frequencies since
oscillations periods are more directly interpreted in terms of
effective particle analogy.

\subsection{Condition of small damping}\label{sec:conditions}

The validity of the small damping approximation for
Eq.~(\ref{eq:effective_equation}) requires
\begin{equation}\label{eq:small_dissipation_condition general}
\Delta \ll \varepsilon_{tot} \ .
\end{equation}
We can make a rough estimate by noting that in the regime of $h \sim
\omega_a$ the value of $R$ is of the order of one. The current
magnitude is assumed to be such that the spin transfer term in
$\alpha_{eff}$ is of the same order as bare $\alpha$. Then the
formula (\ref{eq:integrals_for_DeltaE_T}) gives
$$
\Delta \sim \alpha \Omega \ .
$$
As for the total energy, we assume that it has the same order of
magnitude as the magnetic energy
$$
\varepsilon_{tot} \sim \varepsilon \sim \omega_a \ .
$$
With the above estimates the small damping condition
(\ref{eq:small_dissipation_condition general}) transforms into
\begin{equation}\label{eq:small_dissipation_condition planar}
\alpha \ll \sqrt{\frac{\omega_a}{\omega_p}} \ .
\end{equation}
This is a well known condition of small Gilbert damping in planar
systems.\cite{bazaliy:2007:APL, weinan-e}

We can now show that the current-dependent term in $\tilde\omega
=\omega_a + u^2/\omega_p$ is a small correction to $\omega_a$
whenever the conditions (\ref{eq:small_dissipation_condition
planar}) holds. Indeed, for currents of the same order as the
critical current we have $u \sim \alpha\omega_p$, so
$$
\frac{u^2}{\omega_p} \sim \alpha^2 \omega_p \ll \omega_a \ .
$$
This estimate justifies our statement in
Sec.~\ref{sec:small_damping}.

\section{Comparison with the no-approximation numeric results}
Oscillation periods can be found numerically by solving the LLG
equation without approximations. In practice this means rewriting
the vector equation (\ref{eq:vector_LLG}) in polar angles $\theta$
and $\phi$ of the unit vector ${\bf n} = \{ n_x, n_y, n_z \} =
\{\sin\theta\cos\phi, \sin\theta\sin\phi, \cos\theta \}$. One
obtains a system
\begin{eqnarray}
 \nonumber
\dot\phi &=& \frac{F_{\phi} + \alpha
F_{\theta}}{(1+\alpha^2)\sin\theta} \ ,
 \\
\dot\theta &=& \frac{F_{\theta} - \alpha F_{\phi}}{1+\alpha^2} \ ,
\end{eqnarray}
with
\begin{eqnarray*}
F_{\phi} &=& -\sin\theta\cos\theta (\omega_p + \omega_a\cos^2\phi) -
h\cos\theta\cos\phi
 \\
 &&- u\sin\phi \ ,
 \\
F_{\theta} &=& -\omega_a \sin\theta\sin\phi\cos\phi - h \sin\phi
  + u \cos\theta\cos\phi \ .
\end{eqnarray*}
The system is solved numerically with {\em Mathematica}. When an
oscillating solution exists, the motion of $\bf n$ approaches the
same precession cycle regardless of the initial conditions, as long
as they are its basin of attraction. The period of the precession
motion is measured after the stationary regime is achieved.

Fig.~\ref{fig:compare_large_h} compares the numeric LLG results with
the planar approximation formula for the high field case, $h >
\omega_a$. One observes a very good correspondence. Near the
critical current $u = -u_2$ the periods of oscillations become
infinite and their frequencies drop to zero.\cite{bertotti:2005}
This is easy to understand from the effective particle analogy: at
the critical current the particle travels between the two maxima of
the energy profile that have the same height with the particle
energy being equal to the potential energy at the maximum point. As
it usually happens in such cases, the motion near the maximum is
infinitely slow and the period is infinite.

In the low field case, $0< h < \omega_a$, the comparison of numeric
and approximate results is shown in Fig.~\ref{fig:compare_small_h}.
In the top panel (Fig.~\ref{fig:compare_small_h}A) the Gilbert
damping is set to $\alpha = 0.001$, and the condition
$|\alpha_{eff}| \ll \sqrt{\omega_a/\omega_p}$ is well satisfied for
all current magnitudes on the graph. The correspondence between the
numeric LLG results and the small damping approximation to the
planar equation is very good. As expected, the oscillations period
diverges at $u = -u_2$ and $u = -u_3$, with no oscillations between
the two thresholds.

Fig.~\ref{fig:compare_small_h}B shows results for $\alpha = 0.01$.
We observe good correspondence between the LLG numeric results and
our theory for the small oscillations, but the full rotations regime
shows appreciable differences which become very large near the
critical current. Their origin is the breakdown of the small damping
approximation:  for $\alpha = 0.01$ the currents in the full
rotation regime are so large that the strong inequality
$|\alpha_{eff}| \ll \sqrt{\omega_a/\omega_p}$ is not well satisfied.
To prove that the small damping approximation is the source of the
discrepancy we have solved the planar equation
(\ref{eq:effective_equation}) numerically. The results (red (gray)
dots in Fig.~\ref{fig:compare_small_h}B) correspond very well to
those obtained directly from LLG (black dots in
Fig.~\ref{fig:compare_small_h}B).

It is also instructive to compare the cases of large and small
fields with the same value of Gilbert damping $\alpha = 0.01$
(Fig.~\ref{fig:compare_large_h} and
Fig.~\ref{fig:compare_small_h}B). One can see that the validity
region of the small damping approximation extends at least to $u
\approx -2$ in the case of large field. At the same time in the case
of small fields this approximation is visibly violated for $u
\approx -2$. The relative fragility of the small damping
approximation in the full rotation regime in small fields can be
traced to the presence of two energy maxima, $\phi = \pm \phi_m$,
instead of just one maximum, $\phi = \pi$, in the large fields. The
situation calls for more work on the approximate solutions of the
effective planar equation (\ref{eq:effective_equation}) with
variable damping.

\section{Conclusions}
We have shown that the planar approximation \cite{bazaliy:2007:APL,
bazaliy:2007:PRB} gives good results for the frequencies of spin
transfer oscillators with dominating easy plane anisotropy. Analytic
expressions for the oscillation periods were derived in the limit of
small Gilbert damping. The mechanical analogy, associated with the
effective planar equation, provides a qualitative understanding of
different precession regimes and naturally explains the singular
behavior of the precession frequency near the transition between
in-plane and out-of-plane precessions.

An important advantage of the planar approximation is the fact that
the problem of finding the unperturbed trajectory is drastically
simplified. All one-dimensional trajectories are straight lines
completely characterized by their endpoints. This property should
help to develop theories of the large-angle precession regimes of
planar spin torque oscillators in the presence of temperature
fluctuations or other noise sources.

\section{Acknowledgements}
This work was supported by the NSF grant DMR-0847159.

\end{document}